\documentclass[aps,prc,preprint,tightenlines,showpacs,%
amssymb,byrevtex,superscriptaddress,nofootinbib]{revtex4}  

\usepackage{epsfig,bm,dcolumn}

\begin{document}

\preprint{hep-ph/0311054}


\title{$\Theta^+$ baryon production in $KN$ and $NN$ reactions}


\author{Yongseok Oh}%
\email{yoh@phya.yonsei.ac.kr}

\affiliation{Institute of Physics and Applied Physics,
Yonsei University, Seoul 120-749, Korea}

\affiliation{Thomas Jefferson National Accelerator Facility, 12000
Jefferson Avenue, Newport News, VA 23606}

\author{Hungchong Kim}%
\email{hung@phya.yonsei.ac.kr}

\author{Su Houng Lee}%
\email{suhoung@phya.yonsei.ac.kr}

\affiliation{Institute of Physics and Applied Physics,
Yonsei University, Seoul 120-749, Korea}

\date{\today}


\begin{abstract}

We study $\Theta^+(1540)$ productions in kaon-nucleon ($KN$) and
nucleon-nucleon ($NN$) interactions by assuming that the $\Theta^+$ is
an isosinglet with $J^P = \frac12^+$.
Possible $t$-channel diagrams with $K^*$ exchange are considered in both
reactions as well as $K$ exchange in $NN$ reaction.
The cross section for $np \to \Lambda^0 \Theta^+$, which has not
been considered in previous calculations, is found to be about a factor
of 5 larger than that for $np \to \Sigma^0 \Theta^+$ due to the large
coupling of $KN\Lambda$ interaction.
The cross sections are obtained by setting $g_{KN\Theta}^{}=1$ and
varying the ratio of $g_{K^*N\Theta}^{}/g_{KN\Theta}^{}$ so that
future experimental data can be used to estimate these couplings.
We also find that the isospin relations hold for these reactions.

\end{abstract}


\maketitle

\section{Introduction}

The first experimental observation of a pentaquark baryon, the
$\Theta^+(1540)$, was made in photon-nucleus reaction \cite{LEPS03}.
Later, the existence of the $\Theta^+$ was confirmed by the analyses on
kaon-nucleus \cite{DIANA03}, photon-deuteron \cite{CLAS03-b},
photon-proton \cite{SAPHIR03,KS03}, and neutrino-nucleon reactions
\cite{ADK03}.
Till now, only the upper bound of the $\Theta^+(1540)$ decay width
is known, it is around 9-25 MeV.
Because of its positive strangeness, the minimal quark content of the
$\Theta^+(1540)$ is $uudd\bar{s}$ and thus has hypercharge $Y=2$.
This means that the $\Theta^+$ cannot be a three-quark state, and hence
should be an exotic \cite{Golo71}.
Such a narrow pentaquark state was predicted by chiral soliton model
\cite{DPP97,PSTCG00}.
There, the $\Theta^+$ is an isosinglet and forms a baryon anti-decuplet
with other pentaquark states, which is also anticipated in the Skyrme
model \cite{Chem85,Pras87,Weig98}.
If we consider baryons consisting of four quarks and one anti-quark, the
flavor SU(3) group structure says that such systems can form the
multiplets of $\bm{35}$, $\bm{27}$, $\overline{\bm{10}}$, $\bm{10}$,
$\bm{8}$, and $\bm{1}$.
If the $\Theta^+(1540)$ is an isosinglet, then it would be a member of
the baryon anti-decuplet.%
\footnote{The $\bm{35}$-plet contains isotensor $\Theta$ which has
$I=2$, while the $I=1$ isovector $\Theta$ is a member of the
$\bm{27}$-plet.}
The recently observed $\Xi^{*--}(1862)$, which carries $S=-2$ and
$Q=-2e$ \cite{NA49-03}, could be a member of the anti-decuplet
as its minimal quark content is $dd\bar{u}ss$.
Such a state was also predicted by quark models and soliton models as
an $I=3/2$ member of the baryon anti-decuplet.
Therefore, if confirmed by other experiments, the observation of the
$\Xi^*(1862)$ strongly supports the existence of baryon anti-decuplet
with the {\em isoscalar\/} $\Theta^+$.
However the $I=1/2$ and $I=1$ members of the baryon anti-decuplet are
under debate, since these are crypto-exotic states and cannot be
distinguished from three-quark baryons by the quantum numbers.
Therefore, identifying those members is strongly dependent on the
structure of the low-lying pentaquark states.
In Ref. \cite{JW03}, Jaffe and Wilczek suggested a
diquark-diquark-antiquark picture for the pentaquark anti-decuplet.
For physical states of the $I=1/2$ and $I=1$ baryons, they considered
mixing with pentaquark octet and identified $N(1440)$ and $N(1710)$ as
pentaquark states.%
\footnote{Recently, in a coupled channel model for $\pi N$ scattering,
the J{\"u}lich group claimed that the Roper $N(1440)$ might be a
quasi-bound $\sigma N$ state instead of a genuine three-quark state
\cite{SHSD98-KHKS00}.}
Further extensions of this picture for pentaquark baryons can be
found e.g., in Refs. \cite{OKL03b,DP03b,SZ03,BGS03}.
Other theoretical investigations on the $\Theta^+$ and/or baryon
anti-decuplet can be found, e.g., in
Refs.~\cite{SR03,CCKN03a,CCKN03b,Cheung03,Gloz03a,KL03a,WK03,Pras03,%
JM03,BFK03,IKOR03,Zhu03,MNNRL03,SDO03,Hosaka03,Cohen03-CL03,CFKK03,%
Sasaki03,KK03,HZYZ03,NSTV03}.

The production of the $\Theta^+$ can also be investigated in heavy-ion
collisions as discussed in Refs.~\cite{Rand03,CGKLL03,LTSR03}.
In Ref. \cite{Rand03}, a statistical model is used to predict that the
$\Theta^+$ yield is about 12-14\% of the $\Lambda$ yield in heavy-ion
collisions.
The dependence of the yield on the collision energy was discussed in
Ref.~\cite{LTSR03}.
Moreover, in Ref.~\cite{CGKLL03}, it was claimed that $\Theta^+$ production
can be a useful probe of the initially produced quark-gluon plasma state,
because the number of the $\Theta^+$ formed in the quark-gluon plasma can be
non-trivial and the final state interaction through the hadronic phase is
not large.
However, as emphasized in Ref.~\cite{CGKLL03}, such claims depend crucially
on understanding the strength of the hadronic interactions of the $\Theta^+$.
Thus investigating the hadronic interaction of the $\Theta^+$ is important
not only in understanding its structure but also in probing heavy-ion
collisions.

The elementary production processes of the $\Theta^+$ baryon have been
investigated by several groups.
In Refs. \cite{LK03a,LK03b}, Lin and Ko estimated the total cross sections
of $\Theta^+$ production in photon-nucleon, meson-nucleon, and
nucleon-nucleon reactions.
It is further improved by including the anomalous magnetic moment interaction
terms for $\gamma n$ reaction in Ref.~\cite{NHK03}.
In Ref.~\cite{OKL03a}, we have reported the total and the differential cross
sections for $\gamma p$, $\gamma n$, and $\pi N$ reactions, which was an
improvement over previous studies in that we consistently included $K^*$
exchanges.
Recently, Liu {\it et al.\/} considered the reaction of $\gamma p \to
\pi^+ K^- \Theta^+$ \cite{LKK03}.
Some polarization observables in $\Theta^+$ photoproduction are also
estimated in Ref.~\cite{ZA03}.
In this paper, as a continuation of our efforts to understand the
$\Theta^+$ production processes, we investigate its production in $KN$
and $NN$ reactions.
These reactions were previously investigated in Ref. \cite{LK03a} for
$KN \to \pi \Theta^+$ and $pp \to \Sigma^+ \Theta^+$ reactions, and then the
cross sections were isospin-averaged.
We improve this calculation by considering the $K^*$ exchanges and we also
consider $np \to \Lambda\Theta$ reaction, which will be shown to have larger
cross section than $np \to \Sigma\Theta$.
In addition, we derive isospin relations for different isospin channels in
$KN$ and $NN$ reactions.

As shown in Refs. \cite{NHK03,OKL03a,LKK03,ZA03,HHO03}, the production
cross sections are strongly dependent on the quantum numbers of the
$\Theta^+$, which is an important issue to be resolved.%
\footnote{In Refs.~\cite{NHK03,OKL03a}, it is claimed that the magnitudes
of the cross sections for $\gamma N \to K \Theta$ are strongly dependent
on the parity of the $\Theta^+$.
The cross sections for odd-parity $\Theta^+$ were shown to be much smaller
than those for even-parity $\Theta^+$ by an order of magnitude.
It is also shown that the differential cross section would have
different angular distribution for different parity of the $\Theta^+$
depending on the coupling $g_{K^*N\Theta}^{}$.}
First, the spin of the $\Theta^+$ is believed to be $\frac12$ \cite{JW03}.
But the parity of the $\Theta^+$ is still under debate.
If we assume that every quark is in the S-wave ground state as in the
usual three-quark baryons, the parity of the pentaquark ground state
would be odd because of the existence of one antiquark.
Some quark models \cite{CCKN03a,HZYZ03}, QCD sum rules \cite{Zhu03,SDO03},
and lattice QCD \cite{CFKK03,Sasaki03} support $J^P=\frac12^-$.
However it is also claimed that the state with antisymmetric
spatial wavefunction should be the ground state \cite{JW03}, which
is consistent with the soliton model predictions \cite{DPP97}.
Furthermore, recent quark model studies show that the ground state is in
a P-wave if one includes the orbital motion of the quarks, which makes the
ground state have even parity \cite{CCKN03b,GK03a-GK03b}.
This is also consistent with the heavy pentaquark ($\Theta_c$ and $\Theta_b$)
study in the Skyrme model, which predicts that the ground state has
$J^P = \frac12^+$ while the state with $\frac12^-$ is the first excited
state \cite{OPM94b-OPM94c-OP95}.
Thus, in this paper, we assume that the $\Theta^+(1540)$ has $I=0$ and
$J^P=\frac12^+$.

This paper is organized as follows.
In Sect. II, we compute the total and differential cross sections for
$KN \to \pi\Theta$ reaction.
The cross sections for $NN \to Y\Theta$ is then obtained in Sec. III,
with $Y= \Lambda^0$ and $\Sigma$.
We shall find that the $\Lambda \Theta$ channel has larger cross
sections than the $\Sigma\Theta$ channel by about a factor of 5.
Section IV contains a summary and discussion.

\section{$\bm{KN \to \pi\Theta^+}$ reaction}

\begin{figure}[t]
\centering
\epsfig{file=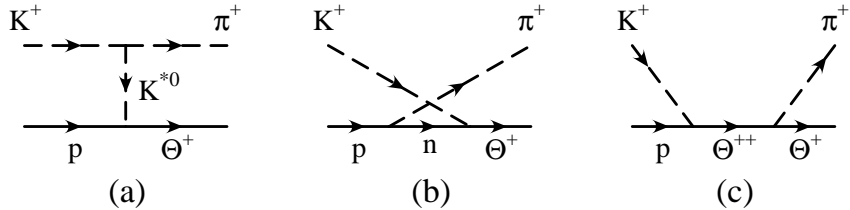, width=0.7\hsize}
\caption{Tree diagrams for $K^+ p \to \pi^+ \Theta^+$ reaction.}
\label{fig:KN}
\end{figure}

As we have discussed before, we assume that the $\Theta^+$ is a $J^P
= \frac12^+$ isoscalar particle belonging to baryon anti-decuplet.
Then in the case of $KN$ reaction, we have four possible isospin
channels in $\Theta^+$ production,
\begin{eqnarray}
&& K^+ p \to \Theta^+ \pi^+, \qquad K^0 p \to \Theta^+ \pi^0, \nonumber
\\
&& K^+ n \to \Theta^+ \pi^0, \qquad K^0 n \to \Theta^+ \pi^-.
\end{eqnarray}
The possible tree diagrams for $K^+ p \to \pi^+ \Theta^+$ are shown in
Fig.~\ref{fig:KN}.
Here we denote the momenta of the incoming kaon, outgoing pion, initial
nucleon, and the final $\Theta^+$ as $k$, $q$, $p$, and $p'$, respectively.
The Feynman diagrams for the other isospin channels can be obtained in the
same way.

There are a few comments regarding the production mechanisms.
First, in Ref. \cite{LK03a}, the authors considered the diagram of
Fig.~\ref{fig:KN}(b).
In this work, however, we extend the model of Ref. \cite{LK03a} by
including $t$-channel $K^*$ exchange, which allows the diagram of
Fig.~\ref{fig:KN}(a).
As we shall see later, the contribution from the $K^*$ exchange is
non-trivial although its magnitude depends on the unknown coupling
$g_{K^*N\Theta}^{}$.
A recent estimate on this coupling gives
$g_{K^*N\Theta}/g_{KN\Theta} \sim 0.6$ \cite{LKK03}.
If this is true, then the contribution from the $K^*$ exchange should be
important.
Second, one may consider other nucleon or $\Delta$ excitations as an
intermediate baryon state in Fig.~\ref{fig:KN}(b).
To begin with, the $\Delta$ excitation is excluded because of its isospin
if the $\Theta^+$ is an isosinglet.
Furthermore, SU(3) symmetry does not allow the coupling of the anti-decuplet
baryon with baryon decuplet and meson octet \cite{OKL03b}.%
\footnote{If the observed $\Theta^+(1540)$ is an isotensor particle
\cite{CPR03}, it should be a member of the {\bf 35} multiplet \cite{Page03}.
Then it has very different selection rules \cite{HL64} and
only the diagram of Fig.~\ref{fig:KN}(b) with the $\Delta$ excitation
as an intermediate state is allowed for its production from $KN$
reactions with two-body final state. Diagrams like Figs.~\ref{fig:KN}(a,c)
are forbidden by isospin in this case.}
Other nucleon resonances such as the nucleon analog of the $\Theta^+$ in the
anti-decuplet can contribute through the diagram of Fig.
\ref{fig:KN}(b).
However, this brings in additional unknown coupling constant for
antidecuplet-antidecuplet-octet interaction.
Thus it will not be considered in this exploratory study.
In Fig. \ref{fig:KN}(c), we have $s$-channel diagram which contains
$\Theta^{++}$ as an intermediate state.
If the $\Theta^{++}$ is an isovector particle, it can contribute to the
production process.
But if it is an isotensor particle, it cannot.
Since the nature, mass,  and couplings of the $\Theta^{++}$ are very
unclear, we do not consider Fig.~\ref{fig:KN}(c) in this paper.

We start with the SU(3) symmetric Lagrangian for the interactions of
baryon anti-decuplet with meson octet and baryon octet \cite{OKL03b},
\begin{equation}
\mathcal{L}_{\overline{D}PB} = -ig \bar{T}^{jkl} \gamma_5
P^j_m B^k_n \epsilon^{lmn} + \mbox{H.c.},
\end{equation}
where $T^{ijk}$ is the baryon anti-decuplet, $P^j_m$ the pseudoscalar meson
octet, and $B^k_n$ the baryon octet.
This leads to
\begin{eqnarray}
\mathcal{L}_{KN\Theta} &=&
 -ig_{KN\Theta}^{} \bar{\Theta} \gamma_5 \bar{K}^c N + \mbox{ H.c.}
\nonumber \\ &=&
 -ig_{KN\Theta}^{} \left( \bar{\Theta} \gamma_5
K^+ n - \bar{\Theta} \gamma_5 K^0 p \right) + \mbox{ H.c.},
\label{KNT}
\end{eqnarray}
where
\begin{equation}
N = \left( \begin{array}{c} p \\ n \end{array} \right), \qquad
K^c = \left( \begin{array}{c} -\bar{K}^0 \\ K^- \end{array} \right).
\end{equation}
The coupling $g_{KN\Theta}^{}$ is related to the universal coupling
constant $g$ by $g_{KN\Theta}^{} = \sqrt{6} g$.
The effective Lagrangian for $K^* N \Theta$ interaction can be obtained
in the same way,
\begin{equation}
\mathcal{L}_{K^*N\Theta} = -g_{K^*N\Theta}^{} \left( \bar{\Theta}
\gamma_\mu
K^{*+\mu} n - \bar{\Theta} \gamma_\mu K^{*0\mu} p \right) + \mbox{
H.c.},
\label{K*NT}
\end{equation}
where we have dropped tensor coupling terms as in Ref.~\cite{OKL03a}.

The other effective Lagrangians necessary for $KN \to \pi N$ reaction are
\begin{eqnarray}
\mathcal{L}_{\pi NN} &=& \frac{g_{\pi NN}^{}}{2M_N} \bar{N} \gamma^\mu
\gamma_5 \partial_\mu \pi N,
\nonumber \\
\mathcal{L}_{K^*K\pi} &=& -i g_{K^* K \pi}^{} \left( \bar{K}
\partial^\mu \pi K_\mu^* - \partial^\mu \bar{K} \pi K_\mu^* \right) +
\mbox{ H.c.}.
\label{eq:lag1}
\end{eqnarray}
Here, we follow the prescription, e.g., of Ref.~\cite{PM02c}, namely, we use
pseudovector coupling for pion interactions and pseudoscalar coupling
for the interactions involving strangeness.
The production amplitude for Fig.~\ref{fig:KN} is given by
\begin{equation}
\mathcal{M}_{K^+p \to \pi^+\Theta^+} = \bar{u}_\Theta(p') \,\mathcal{M}\,
u_p(p),
\end{equation}
where
\begin{eqnarray}
\mathcal{M}_{K^+p}^{(1a)} &=& \frac{\sqrt2 g_{K^*K\pi}^{} g_{K^*N\Theta}^{}}{
(k-q)^2 - M_{K^*}^2}  \left\{
k\!\!\!/ + q\!\!\!/ - \frac{1}{M_{K^*}^2} (M_K^2 - M_\pi^2) (
k\!\!\!/ - q\!\!\!/) \right\}, \nonumber \\
\nonumber \\
\mathcal{M}_{K^+p}^{(1b)} &=& - \frac{\sqrt2 g_{KN\Theta}^{} g_{\pi NN}^{}}{
2M_N\{(p-q)^2 - M_N^2\}} \left\{
q\!\!\!/ - p\!\!\!/ + M_N \right\} q\!\!\!/.
\end{eqnarray}
Each vertex has a form factor in the form of
\begin{equation}
F(r,M_{\rm ex}) = \frac{\Lambda^4}{\Lambda^4 + (r - M_{\rm ex}^2)^2},
\label{ff}
\end{equation}
where $M_{\rm ex}$ and $r$ are the mass and the momentum squared of the
exchanged particle, respectively.
The value of the cutoff parameter $\Lambda$ will be discussed later.

For the coupling constants, we use the well-known value for
$g_{\pi NN}^{}$ as $g^2_{\pi NN}/(4\pi) = 14.0$.
The $K^*$ decaying into $K\pi$ then yields $g_{K^*K\pi}^{} = 3.28$, which
is close to the SU(3) symmetry value, $3.02$.
The coupling $g_{KN\Theta}^{}$ can be, in principle, determined from the
decay width of $\Theta \to KN$.
However, at this moment,
only its upper bound is known from experiments, 9--25 MeV.
Theoretically, the chiral soliton model predicted 15 MeV in Ref.~\cite{DPP97},
which was later improved to be about 5 MeV \cite{PSTCG00}.
If we assume that the $\Theta^+$ decay width is 5 MeV, then we have
$g_{KN\Theta}^{} = 2.2$ \cite{OKL03a}.
Recent analyses on $KN$ elastic scattering data favor such a small decay
width of the $\Theta^+$ \cite{Nuss03-CN03,HK03} or a smaller decay
width, namely 1 MeV or even less \cite{ASW03}.%
\footnote{No evidence for $\Theta^{++}$ in the existing data for $K^+p$ channel
was also reported in Ref.~\cite{ASW03}.}
There is only one experimental information about the total cross
section of $\Theta^+$ production in $\gamma p$ reaction near threshold
\cite{SAPHIR03}, which, however, should be confirmed by further analyses
\cite{add1}.
Therefore in this paper, we do not try to fix $g_{KN\Theta}^{}$.
Instead, we give the results with $g_{KN\Theta}^{} = 1$ so that future
experimental data can be used to estimate the coupling constants with
our predictions.
We also note that $g_{KN\Theta}^{}=1$ corresponds to
$\Gamma(\Theta) \approx 1.03$ MeV.
We also point out that the $\Theta$ couplings to $K^+n$ and
$K^0p$ have different phase, which differs from Ref.~\cite{DPP97}.
Our convention is consistent with the SU(3) symmetry for the
anti-decuplet \cite{OKL03b} and is crucial to obtain the isospin
relations (\ref{KN:iso}).
Finally for $g_{K^*N\Theta}$, there is no information for this coupling.
In Ref. \cite{OKL03a}, we have mentioned that precise measurements on the
differential cross sections for $\gamma N$ and $\pi N$ reactions can be
used to estimate this coupling.
Recently it was estimated to be $g_{K^*N\Theta}^{}/g_{KN\Theta}^{} \sim 0.6$
based on some theoretical assumptions \cite{LKK03}.
Because of the lack of precise information, however, we treat
$g_{K^*N\Theta}^{}$ as a free parameter and give the results by varying its
value as we did in Ref. \cite{OKL03a}.
It should be also noted that the $K^*N\Theta$ interaction contains tensor
coupling.
The contribution from this term should be examined but will not be
discussed in this qualitative study.

The other isospin channels can also be calculated using the effective
Lagrangians above, and it is straightforward to find the following isospin
relation,%
\footnote{In the reactions of $K^+n$ and $K^0p$, one may consider diagrams
like Fig.~\ref{fig:KN}(c) with the $\Theta^+$ as the intermediate state.
However, these diagrams are not allowed since the $\Theta\Theta\pi$
interaction is prohibited by isospin.}
\begin{equation}
\mathcal{M}_{K^+p} = - \mathcal{M}_{K^0 n} = -\sqrt2 \mathcal{M}_{K^0 p}
= -\sqrt2 \mathcal{M}_{K^+ n}.
\label{KN:iso}
\end{equation}
Note that the different phase between the $\bar{\Theta}K^+ n$ and
$\bar{\Theta} K^0 p$ interactions is essential to have the above relation.
In this paper, we give the results for the $K^+p$ reaction only.
Cross sections for the other isospin channels can then be read from our
result by using the isospin relation above.

\begin{figure}[t]
\centering
\epsfig{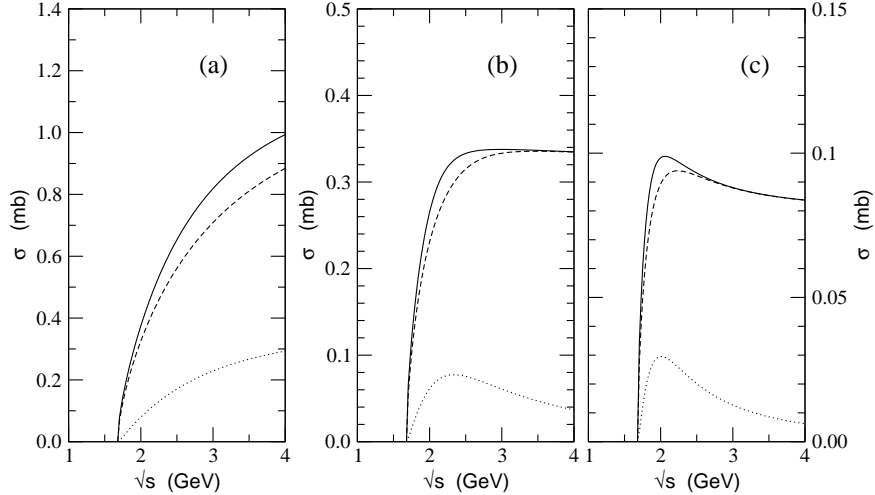}
\caption{Total cross sections for $K^+ p \to \pi^+ \Theta^+$ (a) without
form factor, (b) with form factor (\ref{ff}) and $\Lambda = 1.8$ GeV, (c)
with $\Lambda = 1.2$ GeV. The solid lines are obtained with
$g_{K^*N\Theta}^{} = + g_{KN\Theta}^{}$, the dotted lines are with
$g_{K^*N\Theta}^{} = 0$, and the dashed lines are with $g_{K^*N\Theta}^{}
= - g_{KN\Theta}^{}$.}
\label{fig:kp-tot}
\end{figure}

The total cross section for $K^+p \to \pi^+ \Theta^+$
is  plotted in Fig.~\ref{fig:kp-tot}.
Since we do not have any experimental information, we first present the
result without form factor in Fig.~\ref{fig:kp-tot}(a).
The results with the form factor (\ref{ff}) are then given in
Figs.~\ref{fig:kp-tot}(b,c) with $\Lambda = 1.8$ GeV and 1.2 GeV,
respectively.
The cutoff parameter $\Lambda = 1.8$ GeV is from kaon photoproduction
analyses \cite{JRVD01-JRDV02}, while $\Lambda = 1.2$ GeV is
from $\pi N$ scattering analyses \cite{PM02c}.
Here, the solid lines are obtained with $g_{K^*N\Theta}^{} = +
g_{KN\Theta}$, the dotted lines are with $g_{K^*N\Theta}^{} = 0$, and
the dashed lines are with $g_{K^*N\Theta} = - g_{KN\Theta}^{}$.

\begin{figure}[t]
\centering
\epsfig{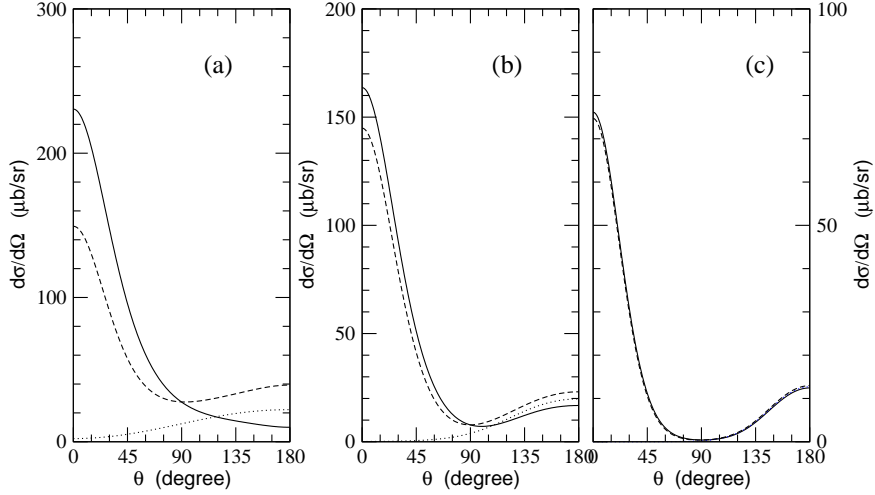}
\caption{Differential cross sections for $K^+ p \to \pi^+ \Theta^+$ at
$\sqrt{s} = 2.4$ GeV (a) without
form factor, (b) with form factor (\ref{ff}) and $\Lambda = 1.8$ GeV, (c)
with $\Lambda = 1.2$ GeV. The dotted line in (c) is almost overlapped with
other lines (so it is not distinguishable) at $\theta \ge 90^\circ$.
It is also suppressed in the other region and is not seen in (c).
The notations are the same as in Fig.~\ref{fig:kp-tot}.}
\label{fig:kp-diff}
\end{figure}

Differential cross section for $K^+p \to \pi^+ \Theta^+$ is shown in
Fig.~\ref{fig:kp-diff} at $\sqrt{s} = 2.4$ GeV as a function of the
scattering angle $\theta$ in the CM frame, where $\theta$ is defined by
the directions of ${\bf k}$ and ${\bf q}$.
Here again, we give the results with different cutoff $\Lambda$.
Figure \ref{fig:kp-diff} shows the role of the $K^*$ exchanges
in a transparent way.
Without $K^*$ exchange, we have the $u$-channel diagram only
[Fig.~\ref{fig:KN}(b)] whose results are given by the dotted lines in
Fig.~\ref{fig:kp-diff}.  They are peaked at backward scatting angles.
With $K^*$ exchanges, the differential cross sections (the dashed and
solid lines) have additional peaks at forward angles.
Thus, the forward peaks in Fig.~\ref{fig:kp-diff} are purely developed
from the $K^*$ exchange and can be larger than the backward peak
depending on $g_{K^*N\Theta}^{}/g_{KN\Theta}^{}$.
So Measurements of differential cross section can give informations
on the magnitude of the coupling $g_{K^*N\Theta}^{}$, which cannot be
estimated from the $\Theta^+$ decay width.
But we find that the phase of $g_{K^*N\Theta}^{}/g_{KN\Theta}^{}$ cannot
be distinguished in these results.

\section{$\bm{NN \to Y \Theta^+}$ reaction}

\begin{figure}[t]
\centering
\epsfig{file=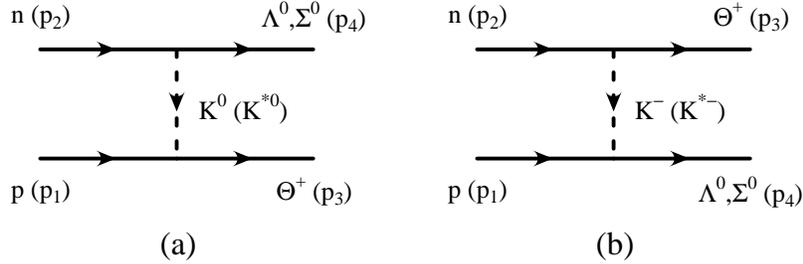, width=0.7\hsize}
\caption{Tree diagrams for $np \to \Sigma^0 (\Lambda^0) \Theta^+ $ reaction.}
\label{fig:NN}
\end{figure}

In this Section, we investigate $NN \to Y\Theta$ reactions where $Y$
stands for $\Sigma$ or $\Lambda$ baryons.
The $pp \to \Sigma^+ \Theta^+$ process was considered within $K$ exchange
model in Ref. \cite{LK03a}.
Here we give more extensive calculation by including $K^*$ exchanges.
We also study $np \to \Lambda^0 \Theta^+$ reaction which was not considered
in Ref. \cite{LK03a}.
To calculate the cross sections, we need additional Lagrangians besides
$\mathcal{L}_{KN\Theta}$ and $\mathcal{L}_{K^*N\Theta}$ given in
Eqs.~(\ref{KNT}) and (\ref{K*NT}), which are
\begin{eqnarray}
\mathcal{L}_{KNY} &=& -i g_{KNY}^{} \bar{N} \gamma_5 Y K
+ \mbox{ H.c.},
\nonumber \\
\mathcal{L}_{K^*NY} &=& - g_{K^*NY}^{} \bar{Y}
\gamma_\mu \bar{K}^{*\mu} N
- \frac{g^T_{K^*NY}}{M_Y + M_N}
\partial^\nu \bar{K}^{*\mu} \bar{Y}
\sigma_{\mu\nu} N + \mbox{ H.c.},
\end{eqnarray}
where $Y = \Lambda,\bm{\Sigma}\cdot\bm{\tau}$.
For $g_{KNY}^{}$, we use the SU(3) symmetry values,
\begin{equation}
g_{KN\Lambda}^{} = - \frac{1}{\sqrt3} g_{\pi NN}^{} (1+2f), \qquad
g_{KN\Sigma}^{} = g_{\pi NN}^{} (1-2f),
\end{equation}
which leads to
\begin{equation}
g_{KN\Lambda}^{}/\sqrt{4\pi} = -3.74 , \qquad
g_{KN\Sigma}^{}/\sqrt{4\pi} = 1.00 ,
\end{equation}
with $d+f=1$ and $f/d=0.575$.
These values are within the range of phenomenological values
\cite{SR99},
\begin{equation}
g_{KN\Lambda}^{}/\sqrt{4\pi} = -4.49 \sim -3.46 , \qquad
g_{KN\Sigma}^{}/\sqrt{4\pi} = 1.32\sim 1.02.
\end{equation}
The $K^*NY$ couplings are estimated in Refs. \cite{SR99,WF88}.
For example, the new Nijmegen potential gives \cite{SR99}
\begin{eqnarray}
&& g_{K^*N\Lambda}^{} = -6.11 \sim -4.26, \qquad
g_{K^*N\Lambda}^T = -14.9 \sim -11.3, \nonumber
\\
&& g_{K^*N\Sigma}^{} = -3.52 \sim -2.46, \qquad
g_{K^*N\Sigma}^T = 4.03 \sim 1.15.
\end{eqnarray}
In our numerical calculation, we use
\begin{equation}
g_{K^*N\Lambda}^{} = -4.26, \qquad
g_{K^*N\Lambda}^T = -11.3, \qquad
g_{K^*N\Sigma}^{} = -2.46, \qquad
g_{K^*N\Sigma}^T = 1.15.
\end{equation}

The transition amplitudes for $np \to \Sigma^0 \Theta^+$ obtained from
Fig. \ref{fig:NN} read
\begin{equation}
\mathcal{M}_{np}^K = -\frac{g_{KN\Sigma}^{} g_{KN\Theta}^{}}{
(p_2-p_4)^2 - M_K^2} \bar{u}(p_4) \gamma_5 u(p_2)\,
\bar{u}(p_3) \gamma_5 u(p_1) + (p_1 \leftrightarrow p_2),
\end{equation}
for $K$ exchange and
\begin{eqnarray}
\mathcal{M}_{np}^{K^*} &=& -\frac{g_{K^*N\Theta}^{}}{(p_2-p_4)^2
- M_{K^*}^2} \left\{ g^{\mu\nu} - \frac{1}{M_{K^*}^2} (p_4-p_2)^\mu
(p_4-p_2)^\nu \right\}
\nonumber \\ && \mbox{} \times
\bar{u}(p_4) \Gamma^{K^*N\Sigma}_\nu(p_4-p_2) u(p_2)\,
\bar{u}(p_3) \gamma_\mu u(p_1)
+ (p_1 \leftrightarrow p_2),
\end{eqnarray}
for $K^*$ exchange with
\begin{equation}
\Gamma^{K^*N\Sigma}_\nu(p_4-p_2) = g_{K^*N\Sigma}^{} \gamma_\nu - i
\frac{g^T_{K^*N\Sigma}}{M_\Sigma + M_N} \sigma_{\nu\alpha}
(p_4-p_2)^\alpha.
\end{equation}
The momenta of the particles are defined in Fig.~\ref{fig:NN}.

The other isospin reactions, $pp \to \Sigma^+\Theta^+$
and $nn \to \Sigma^- \Theta^+$, have the following relation,
\begin{equation}
\mathcal{M}_{pp} = -\mathcal{M}_{nn} = - \sqrt{2} \mathcal{M}_{np}.
\label{eq:NNiso}
\end{equation}
Since the cross sections for $pp \to \Sigma^+\Theta^+$ and $nn \to
\Sigma^- \Theta^+$ reactions can be read off from that for $np \to
\Sigma^0 \Theta^+$ by the above relation, we give the results for $np$
reactions only.

In $np$ reaction, we have an additional channel in the final state,
i.e., $np \to \Lambda^0 \Theta^+$.
The production amplitudes of this reaction read
\begin{eqnarray}
\mathcal{M}_{np \to \Lambda^0 \Theta^+}^K &=& \frac{g_{KN\Lambda}^{}
g_{KN\Theta}^{}}{(p_2-p_4)^2 - M_K^2} \bar{u}(p_4) \gamma_5 u(p_2)\,
\bar{u}(p_3) \gamma_5 u(p_1) + (p_1 \leftrightarrow p_2),
\nonumber \\
\mathcal{M}_{np \to \Lambda^0 \Theta^+}^{K^*} &=&
\frac{g_{K^*N\Theta}^{}}{(p_2-p_4)^2 - M_{K^*}^2} \left\{ g^{\mu\nu} -
\frac{1}{M_{K^*}^2}(p_4-p_2)^\mu (p_4-p_2)^\nu \right\}
\nonumber \\
&& \mbox{} \times
\bar{u}(p_4) \Gamma^{K^*N\Lambda}_\nu(p_4-p_2) u(p_2)\,
\bar{u}(p_3) \gamma_\mu u(p_1)
+ (p_1 \leftrightarrow p_2),
\label{amp:np}
\end{eqnarray}
for $K$ and $K^*$ exchanges, where
\begin{equation}
\Gamma^{K^*N\Lambda}_\nu(p_4-p_2) = g_{K^*N\Lambda}^{} \gamma_\nu - i
\frac{g^T_{K^*N\Lambda}}{M_\Lambda + M_N} \sigma_{\nu\alpha}
(p_4-p_2)^\alpha.
\end{equation}
The form factor (\ref{ff}) is assumed to be multiplied to each vertex.

\begin{figure}[t]
\centering
\epsfig{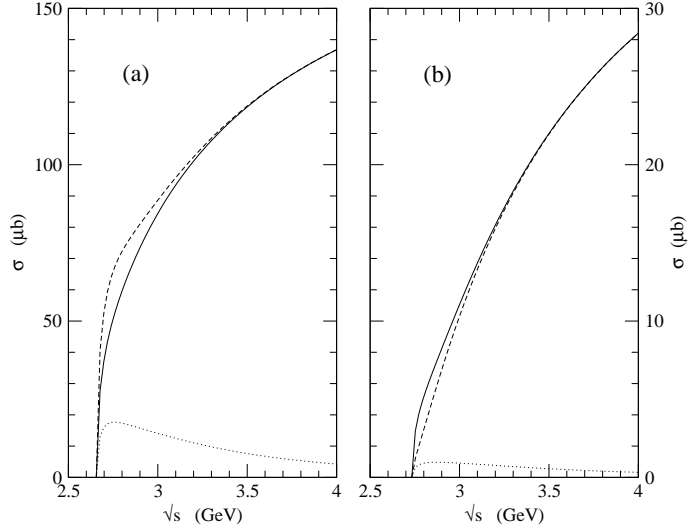}
\caption{Total cross sections for (a) $n p \to \Lambda^0 \Theta^+$ and (b)
$np \to \Sigma^0 \Theta^+$ with the form factor (\ref{ff}) and $\Lambda
= 1.2$ GeV.
The notations are the same as in Fig.~\ref{fig:kp-tot}.}
\label{fig:np-tot2}
\end{figure}

In Fig.~\ref{fig:np-tot1}, the total cross sections for $np \to \Lambda^0
\Theta^+$ and $np \to \Sigma^0 \Theta^+$ are given, which do not include
the form factors.
Shown in Fig.~\ref{fig:np-tot2} are the results with the form factor
(\ref{ff}) and $\Lambda = 1.2$ GeV.
These results show that the $K^*$ exchange dominates the process if
$g_{K^*N\Theta}^{}/g_{KN\Theta}$ is not so small.
They also show that $K$ and $K^*$ exchanges have different energy dependence
in the total cross sections.

\begin{figure}[t]
\centering
\epsfig{file=fig5.eps, width=0.55\hsize}
\caption{Total cross sections for (a) $n p \to \Lambda^0 \Theta^+$ and (b)
$np \to \Sigma^0 \Theta^+$ without form factors.
The notations are the same as in Fig.~\ref{fig:kp-tot}.}
\label{fig:np-tot1}
\end{figure}

\begin{figure}[t]
\centering
\epsfig{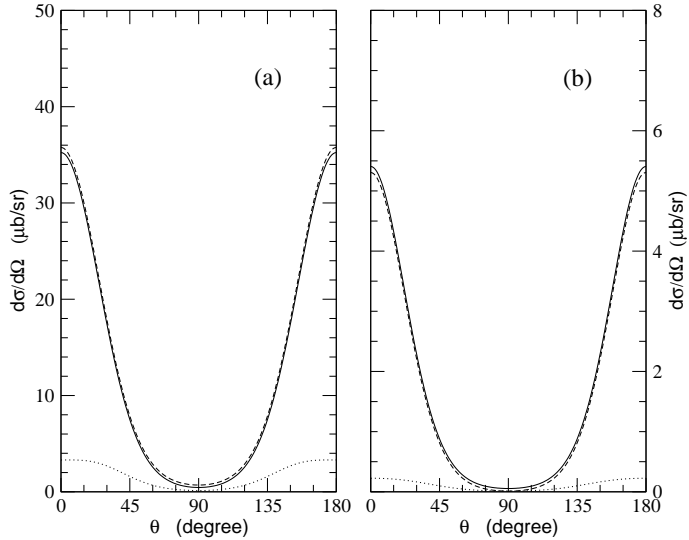}
\caption{Differential cross sections for (a) $n p \to \Lambda^0 \Theta^+$
and (b) $np \to \Sigma^0 \Theta^+$ at $\sqrt{s} = 3$ GeV with the form
factor (\ref{ff}) and $\Lambda = 1.2$ GeV.
The notations are the same as in Fig.~\ref{fig:kp-tot}.}
\label{fig:np-diff2}
\end{figure}

The role of $K^*$ exchange can also be identified in differential cross
sections.
Figures~\ref{fig:np-diff1} and \ref{fig:np-diff2} show the differential
cross sections at $\sqrt{s} = 3$ GeV without and with the form factors
(with $\Lambda = 1.2$ GeV).
The scattering angle $\theta$ is defined by the directions of ${\bf
p}_1$ and ${\bf p}_3$ in CM frame.
It is clearly seen from these figures that the differential cross
sections have symmetric shape about $\theta=90^\circ$ in both
reactions, which can be expected from the structure of the production
amplitudes, e.g., in Eq.~(\ref{amp:np}) regardless of the exchanged
meson.
Since the ratio of the minimum and maximum values of the differential
cross sections depends on the ratio of coupling constants, measurement
of the ratio would shed light on the determination of the magnitude of
$g_{K^*N\Theta}^{}/g_{KN\Theta}^{}$.
But the results are nearly independent on the phase of
$g_{KN\Theta}^{}/g_{K^*N\Theta}$ (Fig.~\ref{fig:np-diff2}).

\begin{figure}[t]
\centering
\epsfig{file=fig7.eps, width=0.55\hsize}
\caption{Differential cross sections for (a) $n p \to \Lambda^0 \Theta^+$
and (b) $np \to \Sigma^0 \Theta^+$ at $\sqrt{s} = 3$ GeV without form factors.
The notations are the same as in Fig.~\ref{fig:kp-tot}.}
\label{fig:np-diff1}
\end{figure}

\section{Summary}

We have estimated the cross sections for $\Theta^+$ baryon production
from $KN$ and $NN$ reactions focusing on the role of the $K^*$ exchanges.
We found that isospin relations hold in $KN \to \pi \Theta$ and
$NN \to \Sigma\Theta$ reactions.
We have also estimated the cross section for $np \to \Lambda^0 \Theta^+$,
which is found to be much larger than that for $np \to \Sigma^0
\Theta^+$.

Without $K^*$ exchange, we found that there is only a backward peak
in the differential cross sections for $KN$ reactions.
The forward peak in Fig.~\ref{fig:kp-diff} is completely ascribed to
the $K^*$ exchange.
Thus precise measurements on the differential cross sections will give a
chance to determine the magnitude of the $g_{K^*N\Theta}^{}$ coupling.
In $NN$ reactions, we found that both $K$ exchange and $K^*$ exchange
give double peaks, forward and  backward peaks, and symmetric
differential cross sections.
Therefore, measuring the ratio of the maximum and minimum values of the
differential cross sections can also give an information on the
couplings.

As we have discussed, investigation of $\Theta^+$ production processes in
meson-nucleon and nucleon-nucleon reactions are useful in understanding
the interactions and the production mechanisms of the $\Theta^+$, as well
as in providing an important information for the $\Theta^+$ yield in
heavy-ion collisions, where hadronic final state effects should be taken
into account.
Experimental studies on these reactions are, therefore, highly required
and might be available at current experimental facilities.

\acknowledgments

Fruitful discussions with V.~Burkert and V.~Kubarovsky are gratefully
acknowledged.
Y.O. is grateful to Thomas Jefferson National Accelerator Facility
for its warm hospitality.
This work was supported by Korea Research Foundation Grant
(KRF-2002-015-CP0074).
The work of Y.O. was supported in part by DOE contract DE-AC05-84ER40150
under which the Southeastern Universities Research Association (SURA)
operates the Thomas Jefferson National Accelerator Facility.

\end{document}